\documentclass[%
 preprint,
 superscriptaddress,
showpacs,
 amsmath,amssymb,
 aps,
 prl,
]{revtex4-1}

\usepackage{color}

\usepackage[utf8]{inputenc}
\usepackage{graphicx}
\usepackage{dcolumn}
\usepackage{bm}

\begin{document}


\title{How to efficiently destroy a network with limited information}
\author{T. M. Vieira} \email{tiagotmv@dfte.ufrn.br}
\affiliation{Departamento de Física, Universidade Federal do Rio Grande do Norte, 
59078-970 Natal, Rio Grande do Norte, Brazil}
\author{G. M. Viswanathan} \email{gandhi@dfte.ufrn.br}
\affiliation{Departamento de Física, Universidade Federal do Rio Grande do Norte, 
59078-970 Natal, Rio Grande do Norte, Brazil}
\affiliation{National Institute of Science and Technology of Complex Systems, 
Universidade Federal do Rio Grande do Norte, 59078-970 Natal, Rio Grande do Norte, Brazil}
\author{L. R. da Silva} \email{luciano@dfte.ufrn.br}
\affiliation{Departamento de Física, Universidade Federal do Rio Grande do Norte, 
59078-970 Natal, Rio Grande do Norte, Brazil}
\affiliation{National Institute of Science and Technology of Complex Systems, 
Universidade Federal do Rio Grande do Norte, 59078-970 Natal, Rio Grande do Norte, Brazil}

\date{\today}

\begin{abstract}
We address the general problem of how best to attack and destroy a
network by node removal, given limited or no prior information about
the edges.  We consider a family of strategies in which nodes are
randomly chosen, but not removed. Instead, a random acquaintance
(i.e., a first neighbour) of the chosen node is removed from the
network.  
By assigning an informal cost to 
the information about the
network structure, we show using cost-benefit analysis that
acquaintance removal is the optimal strategy to destroy networks
efficiently.
\end{abstract}

\maketitle


Networks~\cite{newman-networks, dorogovtsev-networks, dorogovtsev-lectures, 
  cohen-networks} 
have been used todescribe many kinds of systems~\cite{watts-1998, 
  albert-1999, jeong-2000, amaral-2000, liljeros-2001, camacho-2002, girvan-2002,
  newman-networks, dorogovtsev-networks, dorogovtsev-lectures, cohen-networks}.  
In general nodes represent system components and 
edges the interactions between them. 
How the edges are arranged in a networks has great importance because
quantities of interest depend on edge placement~\cite{newman-networks, 
  dorogovtsev-networks, dorogovtsev-lectures, cohen-networks, newman-2008}, e.g.  
connectivity distribution, clustering coefficient, 
resilience to node and edge removal, spreading 
processes, and small-world effects.  The two most studied are networks 
with a typical value of connectivity and scale-free networks. 
In the former the edges are placed between completely random node pairs 
following the Erdös-Rényi (ER) algorithm~\cite{erdos-1960}.  
In the latter the edges are placed randomly with a bias 
towards more connected nodes according to the Barabási-Albert (BA) 
algorithm~\cite{barabasi-1999}. The vulnerability of network to attacks, 
the strategies used in these attacks and the strategies' efficiency have
been extensively studied since 2000s~\cite{callaway-2000, albert-2000,
  albert-2001, holme-2002, cohen-2000, cohen-2001, cohen-2003,
  gallos-2005, gallos-2007, holme-2004, bellingeri-2014}.  A network
can be attacked by removing nodes or edges.  When an edge is removed,
the rest of the network remains unchanged.  When a node is removed, 
standard practice is to remove all the edges linked to the removed node. 
In this paper we study how to destroy ER and BA networks 
through node removal. 
Among the various ways to remove nodes, two basic forms are 
particularly interesting \cite{albert-2000, albert-2001}. The
random strategy consists of removing nodes randomly. 
The targeted strategy consists of removing nodes in order
according to their connectivity, from the highest connected nodes to
the lowest connected.
These two strategies are limiting cases: the random strategy requires
zero information about the edges, whereas the targeted strategy
demands that all edges are known in advance. To target the nodes in
order, one needs to know the full network structure, i.e. one needs
the complete information about the network.

Apart from these two basic strategies, we present a family of 
strategies based on an idea introduced by Cohen \emph{et al.}
\cite{cohen-2003} in a different context: 
the acquaintance immunization strategy.  
The idea is as follows: instead of immunizing a randomly chosen node, 
it's better to immunize a random acquaintance of the node. 
We refer to this as the acquaintance strategy.  
Compared to a single random choice, these two combined random choices 
increase the probability that highly connected nodes are immunized, 
even though their connectivity remains unknown. 
Here, we adapt this idea to the network destruction context. 

\begin{figure}[f]
\includegraphics[width=10cm]{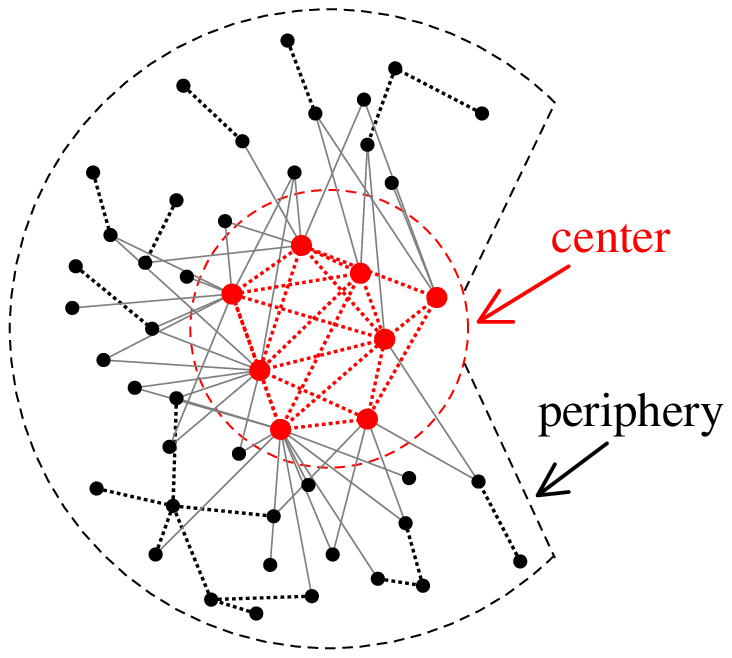}
\caption{\label{fig-network}(Color online) Illustration of network
  center (red) and periphery (the remaining nodes), showing that most
  of the edges connect center and periphery (solid gray edges).
  Informally, the central nodes have much higher connectivity or
  degree.  Independently of the rules governing the distribution of
  the edges in a network, random assignment or preferential
  attachment, the number of nodes in the center is much smaller than
  the number of nodes in the periphery when the total number of nodes
  in the network is very large. Nevertheless, the number of edges
  inside the center (red, dotted) and inside the periphery (black,
  dotted) are typically small compared to the total number of
  edges. The majority of edges (solid gray) connect peripheral nodes
  to central nodes. Key consequences are (i) when a node is randomly
  chosen, it is likely a peripheral node and (ii) when one of its
  acquaintance (or first neighbour) is randomly selected, most likely
  the acquaintance belongs to the center.  These two facts motivate a
  simple strategy to destroy the center: randomly attack the
  acquaintances of random chosen nodes.}
\end{figure}

We first explain in more detail why the acquaintance strategy works.
Fig.~\ref{fig-network} illustrates how most networks can be separated
into two parts (node sets): a ``center'' formed by the most connected
nodes and a ``periphery'' with all remaining nodes.  There are some
exceptions to this rule, such as regular networks, which we ignore
here.  The number of peripheral nodes is greater than the number of
nodes in the center and the difference grows when the total number of
nodes increases. As there are few nodes in the center and each has
many edges, the vast majority of these edges connects central nodes to
peripheral nodes.  Similarly, each peripheral node has few edges and,
on average, only a small portion connects two peripheral nodes. Hence
the majority of edges of a peripheral node have a central node at the
other end. The two combined random choices work together as follows:
(i) the first random choice over all the nodes has greater probability
of choose a peripheral node, and (ii) the second random choice over
this first selected node's acquaintances has greater probability of
picking out a central node. Therefore, the acquaintance strategy is
more easily capable of reaching the central nodes, compared to the
random strategy. It essentially exploits the existence of edges 
connecting the periphery to the center. 

However, because the acquaintance strategy don't consider node connectivity, 
the order in which it removes the nodes is not the best. The strategy 
that implements the best order for removing nodes is the targeted strategy. 
Seeking to upgrade the acquaintance strategy's performance, 
improvements on the original idea have been proposed 
\cite{holme-2004,  gallos-2007}, 
but they require knowledge of the number of edges for a given node. 
We take a different approach and suggest an improvement that indirectly 
exploits the connectivity, but does not require its direct knowledge.  Specifically, 
\emph{we propose that a node must be chosen as an acquaintance 
  more than once before it is removed}.  
The basic insight is that some acquaintances are more central than others.
In other words, the more edges a node has, the more likely it is chosen 
as an acquaintance.

We now impose a threshold $n_r$ for node removal.  
Associate to each node $j$ a memory parameter $n_j$. 
Initially all the nodes have $n_j=0$, but when a node is selected
among the acquaintances of another randomly chosen node, it increments
to $n_j=1$. Again if the same node is selected between the
acquaintances afterwards, it increases to $n_j=2$, and so on.  When it
attains the threshold $n_r$, then the node is removed.  A given node
may be selected from the acquaintances of a same node more than once.
In this approach, the original acquaintance strategy described
previously corresponds to the special case $n_r=1$.  This slight
modification increases the probability that the most connected nodes
are the first to be removed.

\begin{figure}[f]
\includegraphics[width=17cm]{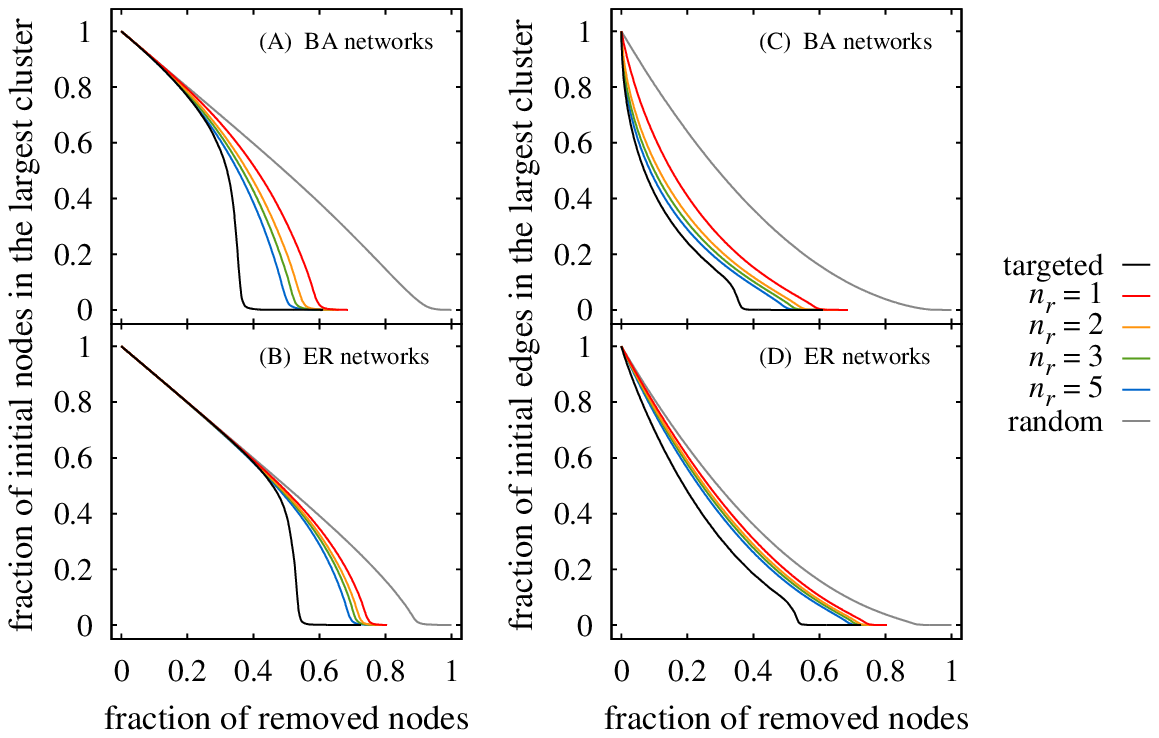}
\caption{\label{plot-01-ab-cd}(Color online) Attacking randomly chosen 
  acquaintances is more destructive than attacking the original 
  randomly chosen nodes.  Plots show the fraction of initial nodes 
  remaining in the largest cluster after attack (plots A and B), 
  and similarly the fraction of remaining initial edges (plots C and D). 
  For all the plots, the leftmost curve corresponds to the targeted 
  strategy, the rightmost corresponds to the random strategy. 
  The four middle curves refer to the family of acquaintance strategies, 
  in which every node has a parameter $n_j$ that counts how many times 
  it has been selected as an acquaintance. Each particular strategy 
  has a defined threshold $n_r$ such that when a node has $n_j=n_r$, 
  it is removed. From left to right: acquaintance curves for $n_r=5$, 
  $n_r=3$, $n_r=2$ and the simple acquaintance case with $n_r=1$. 
  Erd\"os-R\'enyi (ER) networks are more resistant in general than 
  Barabási-Albert (BA) networks. But BA networks are slightly less 
  fragile to the random strategy.
  Remembering that each removed node carries away the edges 
  connected to it, the curves in plots C and D reveal how 
  the removing order of nodes for each strategy is closer to 
  the best order (targeted strategy). So, the faster the 
  curve falls, the more connected were the removed nodes. 
  We consider that the network is destroyed when there are no edges 
  connecting the nodes that haven't been removed yet, so the curves 
  extends only to this point. 
  This same pattern plot is kept for all the other graphs. 
  These results, and all others hereafter, are based on simulation of 100 
  networks of each type and each network has 10000 nodes and 49985 edges. 
}
\end{figure}

The value of the threshold $n_r$ thus generates a whole family of
acquaintance strategies, such that each strategy is characterized by a
specific value for the removal threshold $n_r$. The number of nodes
that must be removed to destroy a network resulting from the
application of all the mentioned strategies is shown in
Fig.~\ref{plot-01-ab-cd}. While plots A and B show how the network's
largest cluster loses nodes when a network is attacked, plots C and D
show how the largest cluster loses edges at the same time.  In
summary, these plots have a clear interpretation: even though
acquaintance strategies are based on only two combined random choices,
they can destroy a network faster than the random strategy. Moreover,
they have better performance on BA rather than ER networks. The plots
also confirm that the memory parameter represents an improvement on
the original acquaintance idea, because the acquaintance curves
deviate from the random strategy curve and approach the targeted 
strategy curve for higher $n_r$.

\begin{figure}[f]
\includegraphics[width=17cm]{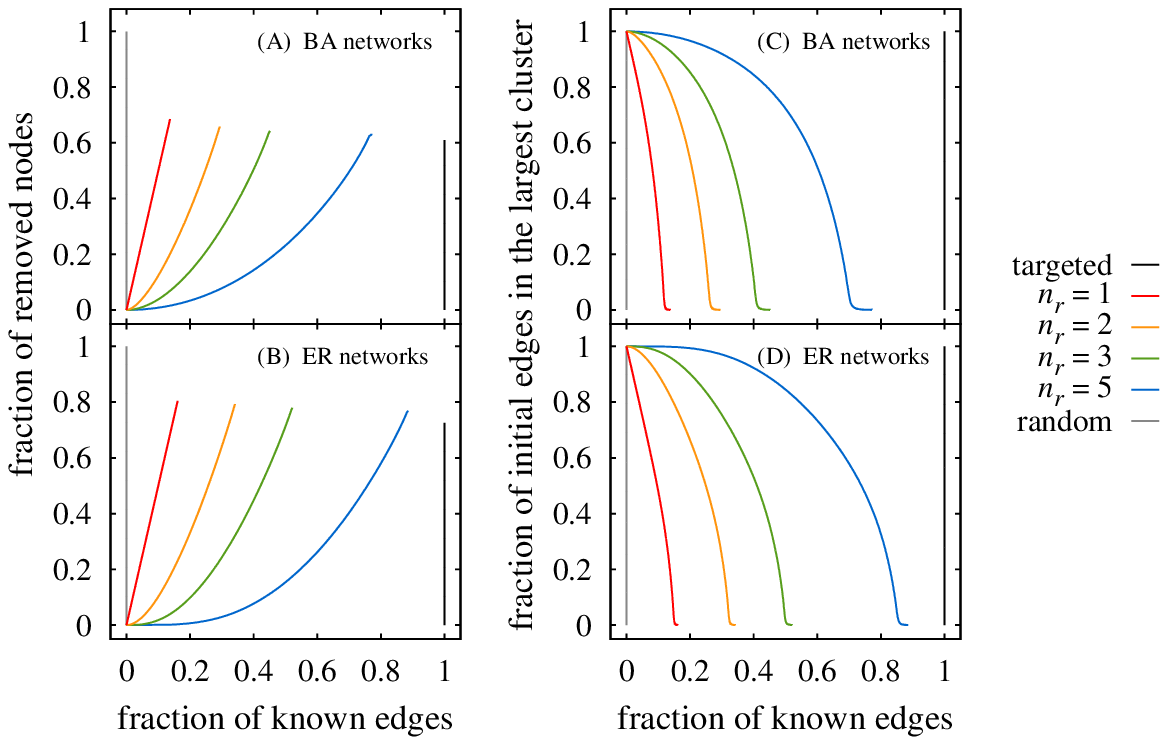}
\caption{\label{plot-02-ab-cd}(Color online) Analysis of the informal
  cost involving in attacking a network following one of the cited
  strategies. Plots show the fraction of removed nodes (plots A and B)
  and the fraction of initial nodes remaining in the largest cluster
  (plots C and D) in relation to the informal cost, that is, the
  number of edges that need to be known at that moment. For all the
  plots, the curves from left to right are: random, $n_r=1$, $n_r=2$,
  $n_r=3$, $n_r=5$ and targeted curves, using the same nomenclature of
  Fig.~\ref{plot-01-ab-cd}.  The targeted strategy (vertical line at
  left) and the random strategy (vertical line at right) are limiting
  cases: the former requires that all the edges be known, while the
  latter can be carried out without any edge be known. Between these
  two there is a family of curves relative to the acquaintance
  strategy. The basic acquaintance removal ($n_r=1$) demands that only 
  a small fraction of the edges be known. This is a demonstration that
  the acquaintance strategy without memory parameter is efficient from
  the viewpoint of the number of edges that must to be accessed before
  a network be destroyed. However by this viewpoint, when the memory
  parameter is considered, more edges need to be known and this aspect
  of the efficiency is progressively lost. It's worth noting that the
  less nodes a strategy removes until destroying a network, the more
  costly it is.}
\end{figure}

Thus far, we have shown the damage to the largest cluster when we
remove a given fraction of nodes (Fig.~\ref{plot-01-ab-cd}).  However,
this is just one aspect of the efficiency.  How much information about
a network do we need access to remove a given fraction of nodes using
each strategy?  Considering that the information about the structure
of a network is contained in the edges, 
\emph{we attributed to knowledge of each edge an informal information cost}.
We consider only networks with simple unweighted edges, so that the
same cost is assigned to every edge. Thus the total cost associated to
the execution of each strategy is given by the number of edges known
during the node removal process.  In Fig.~\ref{plot-02-ab-cd} this
issue is examined and the plots highlight the cost of information: the
random and targeted strategies are limiting cases, the former having
zero cost associated and the latter having maximum possible cost. The
family of acquaintance strategies has a cost proportional to the
characteristic $n_r$ of each particular strategy, so that the
acquaintance strategy without memory is that with minor cost between
them.  In fact, the execution of the original acquaintance idea has a
cost with linear growth during all the removing process, which makes
this strategy optimal if we only take into account the cost-dependent
aspect (because for each edge known, one node is removed).  It's
noteworthy that the acquaintance strategies don't require any prior
information about the networks, they are such that as soon as
information is acquired, it can be immediately used for node
removal.

A complete analysis of efficiency only can be done when the plots
shown in Figs.~\ref{plot-01-ab-cd} and \ref{plot-02-ab-cd} are
considered altogether. Unfortunately a direct way to do this analysis
doesn't exist and we can do it only qualitatively. Indeed, {the
  appropriate strategy depends on the availability of information
  about the network to be destroyed}. If the network can be examined
in all its extension, such that all the information about it is
available, the appropriate strategy is to attack from the most
connected nodes, i.e. the targeted strategy. If only a limited amount
of information can be accessed, then a balance between the number of
edges that can be known and the number of nodes that should be removed
until the network destruction must be considered. In this case one of
the acquaintance strategies is most appropriate.

We comment on a secondary finding.  The fragility of BA
networks to attacks is evident when compared to ER networks, except
for the random strategy. Even in this case, the difference in favour
of BA networks only can be noted when most of the nodes have already
been removed. Consider strategies which aim to destroy networks from
the central nodes. Fig.~\ref{plot-02-ab-cd} shows that the cost
involved in executing each strategy is almost the same for both types
of networks, yet BA networks are highly susceptible to the loss of
edges, according to plots C and D in Fig.~\ref{plot-01-ab-cd}, and
their largest clusters break up faster, according to plots A and B.
This is a consequence of the form in which information about the
structure of the networks, i.e. the edges, is distributed between the
nodes.  This point deserves a further discussion.

When the same strategy is used to attack networks of different types,
the fragilities of each type will determine how efficient the strategy
is.  For example, ER networks don't have a well-defined lower limit
for the connectivity of its nodes allowing that some nodes have only
one edge linked to them, so that the removal of these nodes can break
the networks into fragments. However, these fragments will preserve a
good part of the network because very few edges have been lost with
the removed nodes. Furthermore, it is hard to reach these nodes, since
they have very few neighbours, which allows the random strategy to
reach them more efficiently than a connectivity-driven strategy. In
other words, the information about the network is dispersed over its
entire extent. In contrast, BA networks have a fundamental fragility:
the existence of extremely connected nodes. They are few, but they
concentrate a great number of edges around them. Any information about
the structure of these networks points to the same regions, to the
same nodes. So, the removal of these hubs can weaken the network,
sometimes to the point at which it will quickly become fragmented into
tiny pieces. In this case the information about the network is
concentrated around a few regions, rendering the network more
susceptible to strategies that exploit connectivity, even indirectly, 
such as the acquaintance strategies studied here.

\begin{acknowledgments}
The authors thanks the financial support of CNPq.
\end{acknowledgments}

\bibliography{arxiv-acquaintance-v16}

\end{document}